\documentclass[aps,prl,twocolumn,showpacs,superscriptaddress,floatfix]{revtex4-1}
\usepackage[utf8]{inputenc} 
\usepackage{amssymb,amsmath}
\usepackage{graphicx}
\graphicspath{{Figures/}}
\usepackage{braket}
\usepackage{siunitx}
\usepackage[dvipsnames]{xcolor}
\usepackage[normalem]{ulem}
\usepackage{physics}
\usepackage{hyperref}
\usepackage{floatrow}
\usepackage{etoolbox}
\newtoggle{SUPPORT}
\togglefalse{SUPPORT}

\hypersetup{
    colorlinks=true, 
    linkcolor=blue,  
    citecolor=blue,  
    filecolor=blue,  
    urlcolor=blue    
}

\definecolor{green}{rgb}{0.0,0.5,0.0}
\usepackage[normalem]{ulem}

\newcommand{\edit}[2][]{ {\textcolor{black}{#2}}}
\newcommand{\editthird}[2][]{ {\textcolor{black}{#2}}}
\def\N23{^{23}\mathrm{N}}                    		    

\begin{document}

\title{\edit[]{Accurate Determination of }Hubble Attenuation and Amplification in Expanding and Contracting Cold-Atom Universes}

\author{S.~Banik}
\affiliation{Joint Quantum Institute, National Institute of Standards and Technology, and University of Maryland, College Park, Maryland, 20742, USA}

\author{M.~Gutierrez~Galan}
\affiliation{Joint Quantum Institute, National Institute of Standards and Technology, and University of Maryland, College Park, Maryland, 20742, USA}

\author{H.~Sosa-Martinez}
\affiliation{Joint Quantum Institute, National Institute of Standards and Technology, and University of Maryland, College Park, Maryland, 20742, USA}

\author{M.~Anderson}
\affiliation{Joint Quantum Institute, National Institute of Standards and Technology, and University of Maryland, College Park, Maryland, 20742, USA}

\author{S.~Eckel}
\affiliation{Sensor Sciences Division, National Institute of Standards and Technology, and University of Maryland, Gaithersburg, Maryland, 20899, USA}

\author{I.~B.~Spielman}
\affiliation{Joint Quantum Institute, National Institute of Standards and Technology, and University of Maryland, College Park, Maryland, 20742, USA}

\author{G.~K.~Campbell}
\affiliation{Joint Quantum Institute, National Institute of Standards and Technology, and University of Maryland, College Park, Maryland, 20742, USA}
\email{gretchen.campbell@nist.gov}
\date{\today}

\begin{abstract}
In the expanding universe, relativistic scalar fields are thought to be attenuated by ``Hubble friction'', which results from the dilation of the underlying spacetime metric.
By contrast, in a contracting universe this pseudo-friction would lead to amplification.
Here, we experimentally measure\edit{, with five-fold better accuracy,} both Hubble attenuation and amplification in expanding and contracting toroidally-shaped Bose-Einstein condensates, in which phonons are analogous to cosmological scalar fields.
We find that the observed attenuation or amplification depends on the temporal phase of the phonon field, which is only possible for non-adiabatic dynamics\edit[, in contrast to the expanding universe in its current epoch, which is adiabatic]{}.
The measured strength of the Hubble friction disagrees with recent theory [J. M. Gomez Llorente and J. Plata, {\it Phys. Rev. A} {\bf 100} 043613 (2019) and S. Eckel and T. Jacobson, {\it SciPost Phys.} {\bf 10} 64 (2021)]\edit[, suggesting that our model does not yet capture all relevant physics]{; because our experiment probes physics outside the scope of this theory---with large excitations in rings of intermediate thickness---this indicates the presence of new physics.}
\end{abstract}

\maketitle

During the early universe's rapid expansion, primordially fluctuating scalar fields are thought to have been exponentially redshifted and attenuated by the expanding spacetime metric, where ``Hubble friction'' contributes to the latter~\cite{Baumann2011}.
Unlike true friction, Hubble friction is non-dissipative and therefore, while it attenuates scalar fields in an expanding universe, it would amplify them in a contracting universe.
In previous work \cite{Eckel2018}, our group showed that an atomic Bose-Einstein condensate (BEC) in an expanding toroidal trap could simulate elements of an expanding universe, including the redshifting of phonons in analogy to the redshifting of photons.
Here, we build upon these studies by: including contracting universes; measuring both Hubble attenuation and amplification with five-fold increased precision; and showing that the magnitude of Hubble friction disagrees with recent theoretical work~\cite{Llorente2019,Eckel2020}.

While the study of astrophysical systems is ordinarily limited to observations, the development of well-controlled laboratory systems has enabled tabletop realizations of general relativistic phenomena.
Examples from a variety of physical platforms ranging from classical fluids to cold atomic systems include: the realization of acoustic black hole horizons~\cite{Rousseaux2008,Philbin2008,Nguyen2015}; stimulated and spontaneous Hawking radiation~\cite{Belgiorno2010,Rubino2012,Steinhauer2016}; and scattering processes around rotating black holes~\cite{Torres2017}.
With their unprecedented control and measurement capabilities, ultracold atoms are an emerging platform for realizing minimal models relevant to high energy physics~\cite{Cao2011}, astrophysics~\cite{Unruh1981, Steinhauer2014, Ohashi2020, Kolobov2021}, and cosmology~\cite{Garay2000, Barcelo2003, Fischer2004, Eckel2018}.

In BECs, phonons are scalar fields that evolve approximately according to an effective spacetime metric defined by the background BEC~\cite{Barcelo2003}.
For toroidally-shaped BECs, expanding or contracting 1D universes can be simulated by dynamically changing the BEC's radius and observing the evolution of azimuthal phonons.
Unlike the expansion observed in the photon-dominated epoch of the universe, we explore non-adiabatic expansions and contractions where the rate of the metric change exceeds the oscillation frequency. 

\begin{figure}[b]
    \centering
    \includegraphics[width=8.6cm]{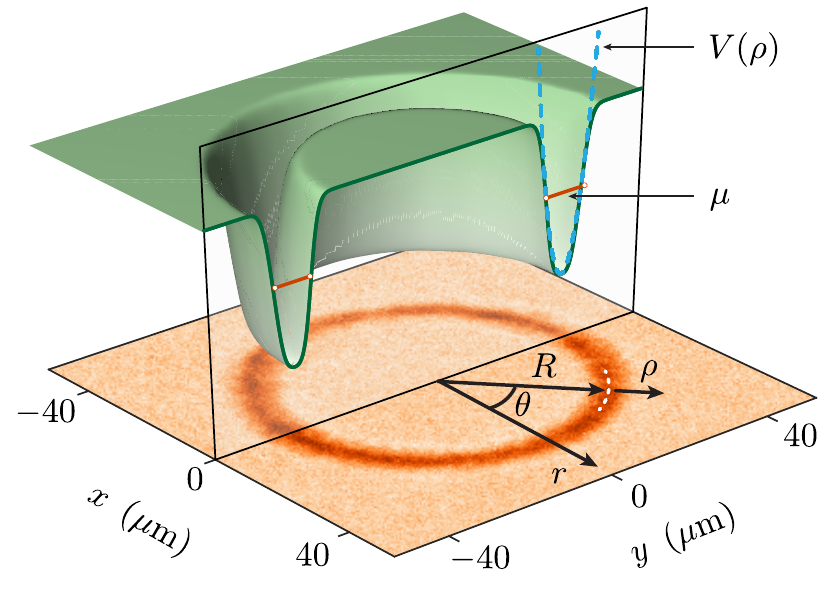}
    \caption{Ring-trapping potential and resulting atomic density. 
    The green surface schematically depicts the trapping potential; the orange lines mark the typical chemical potential $\mu$.
    The blue-dashed curve shows a power law fit to the potential (up to $\mu$) around $\rho=$\edit[]{$|r-R| =$} 0 giving exponent $2.02(3)$ for this example.
    The measured 2D density $n_{\rm 2D}(\rho,\theta)$ is shown in the ${\bf e}_x\!-\!{\bf e}_y$ plane (with peak density $165~\mu\rm{m}^{-2}$) and the white dashed arc marks the mean radius $R$.
    Because of the short $500~\mu{\rm s}$ TOF, the observed width of the ring is slightly in excess of that anticipated from the in-situ T-F approximation.
    }\label{fig:coordinates}
\end{figure}

Phonons are predominately phase excitations with respect to the BEC's order parameter.
For a toroidal BEC with radius $R(t)$ (see Fig.~\ref{fig:coordinates})\edit[]{and 3D volume $\mathcal{V}(t)$}, azimuthal phonons with mode number $m$, have an approximate phase profile $\delta \phi_{\rm 1D} (\theta,t) \equiv \delta \phi(t)\sin(m\theta)$ independent of $r$ and $z$ and obey the wave equation~\cite{Eckel2020}
\begin{align}
    \left\{\partial^{2}_t  +  \left[2\gamma + \frac{\dot{\mathcal{V}}(t)}{\mathcal{V}(t)}\right] \partial_t  + \omega_m^2(t)\right\}\delta \phi(t) &= 0
    \label{eq:dynamics_DE}
\end{align}
at low energy (i.e., small $m$).
Here, the instantaneous angular frequency is $\omega_m = m c_\theta(t) / R(t)$, for speed of sound $c_\theta(t)$.
Because this manuscript focuses exclusively on the $m=1$ mode, we omit the $m$ subscript in what follows.
The quantity in square brackets is reminiscent of damping because it multiplies the first derivative of time.
It includes two terms: a phenomenological damping constant $\gamma$~\footnote{This phenomenological damping term can account for Landau and Beliaev damping mechanisms~\cite{Chiang2009} as well as imperfections in the confining potential.} and the non-dissipative ``Hubble friction'' $\dot{\mathcal{V}} / \mathcal{V}$ arising from the changing metric defined by the background condensate. 
We model the external potential (see Fig.~\ref{fig:coordinates}) as quadratic in $z$ and power law in $\rho=|r-R|$; in the Thomas-Fermi (T-F) and thin-ring approximations, these lead to the BEC's 3D volume $\mathcal{V} \propto R^\alpha$ and speed of sound $c_\theta\propto R^{-\alpha/2}$, where the value of the constant $\alpha$ depends on the potential~\cite{Eckel2020}.
Rather than detecting $\delta \phi_{\rm 1D}$, we measure the associated density perturbation $\delta n_{\rm 1D} (\theta, t)= \delta n(t) \sin(m\theta)$.
The relationship between $\delta \phi$ and $\delta n$ is $\partial_t \delta\phi = -(g/\hbar) (\delta n/R^\alpha)$, in terms of the Gross–Pitaevskii equation~\cite{Dalfovo1999} interaction constant $g$.

\begin{figure}[!t]
    \centering
    \includegraphics[width=8.6cm]{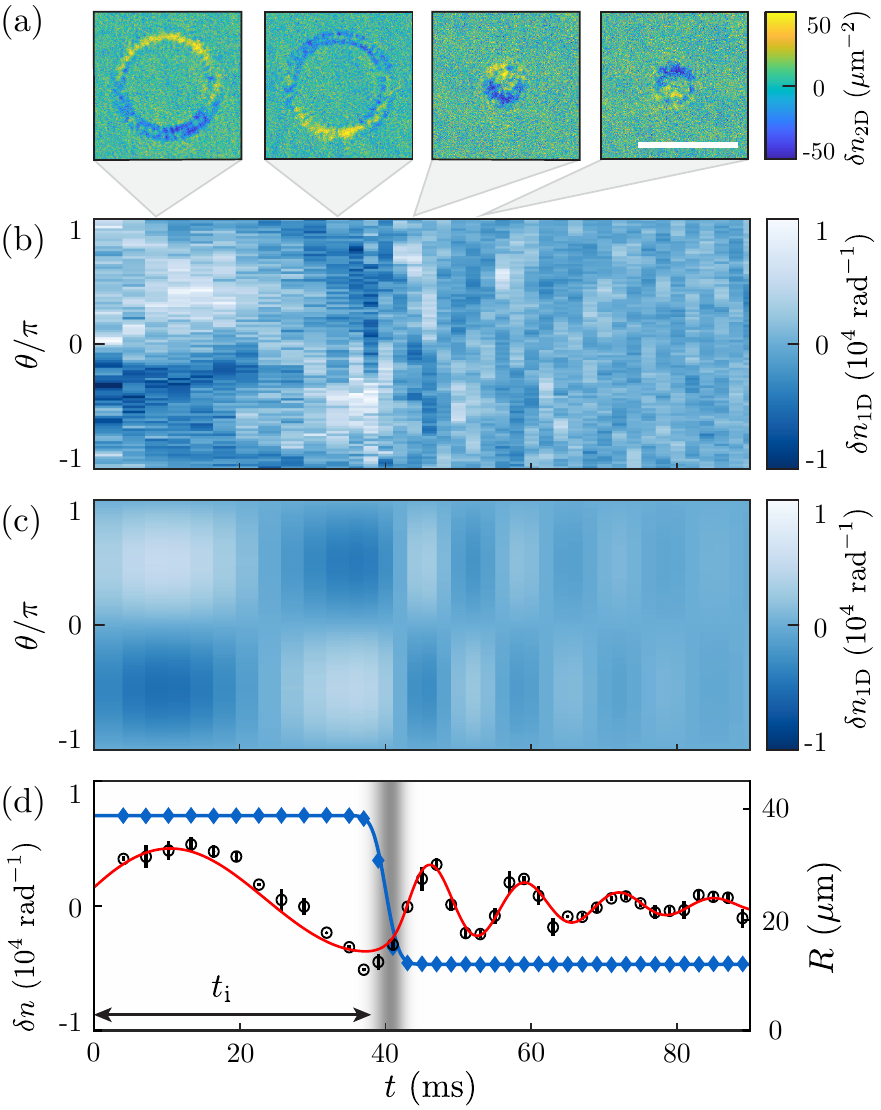}
    \caption{Phonon evolution in a contracting toroidally-shaped BEC, averaged over three measurements.
    (a) Density perturbations for a ring with $R_{\rm i} =  38.4(6)\,\mu{\rm m}$ at 10 ms and 35 ms, and $R_{\rm f} = 11.9(2)\,\mu{\rm m}$ at 45 ms and 53 ms. The density scale of images before contraction is multiplied by 2.
    The horizontal bar corresponds to 80 $\,\mu{\rm m}$.
    (b) Experimental measurements and (c) fit to Eq.~\eqref{eq:dynamics_DE} of angular density perturbation $\delta n_{\rm 1D}$ as a function of azimuthal angle $\theta$ and time $t$, where the ring contraction occurs at $t_{\rm i}$.
    (d) Phonon amplitude $\delta n$ as a function of time.
    The circles plot the phonon amplitude obtained from fitting each time-slice of (b) to a sinusoid~\cite{ErrorFootnote}.
    The red curve is the instantaneous amplitude from the fit in (c).
    The diamonds are the measured mean radius of the BEC and the blue line is the programmed radius of the trap.
    The grayscale bar encodes the value of $|\dot{R}/R|$, with a maximum of $328(11)~{\rm s}^{-1}$ at $t_{\rm peak} = 41~{\rm ms}$.
    The arrow indicates $t_{\rm i} = 38.2\,{\rm ms}$.
    }\label{fig:ods}
\end{figure}

In our experiments, the potential $V(\rho)$ is nominally fixed during expansion or contraction, predicting $\dot{\mathcal{V}}/\mathcal{V} = \gamma_{\rm H} \dot{R}/R$ with strength $\gamma_{\rm H}=\alpha$.
In expanding systems ($\dot{R}>0$) the Hubble friction term attenuates phonons, while in contracting systems it amplifies them.
In the non-adiabatic regime $\dot{R}/R\gtrsim \omega_m$, the timing of expansion or contraction relative to the phonon's temporal phase becomes important for subsequent dynamics.
We show this enhances or diminishes the impact of Hubble friction: because the Hubble friction term includes the product of $\dot{R}/R$ and $\delta n(t) \propto \partial_t \delta \phi(t)$, tuning the timing of expansion or contraction relative to the oscillation changes the degree of amplification or attenuation.

Our experiments~\cite{Lin2009,Kumar2016} begin with quasi-2D \edit[]{$^{23}$Na} BECs with $N\approx 1\times10^5$ atoms confined in a pair of blue-detuned ($\lambda = 532\ {\rm nm}$) optical dipole traps.
The chemical potential is $\mu \approx h\times 2.7\ {\rm kHz}$.
The harmonic vertical confinement, with frequency $\omega_z/2\pi \approx 1.2\ {\rm kHz}$, is provided by a horizontally propagating Hermite-Gauss ${\rm TEM}_{01}$ beam.
We generate nearly arbitrary space and time-dependent potentials in the $r$-$\theta$ plane by imaging $\lambda = 532\ {\rm nm}$ laser light reflected by a digital micro-mirror device (DMD) onto the BEC.
We use these potentials to create toroidal traps with radius $R$ (see Fig.~\ref{fig:coordinates}) ranging from $12\ \mu{\rm m}$ to $39 \ \mu{\rm m}$ and \edit[]{projected }radial width $\approx 5\ \mu{\rm m}$.

\begin{figure*}[t!]
    \centering
    \includegraphics[width=17.8cm]{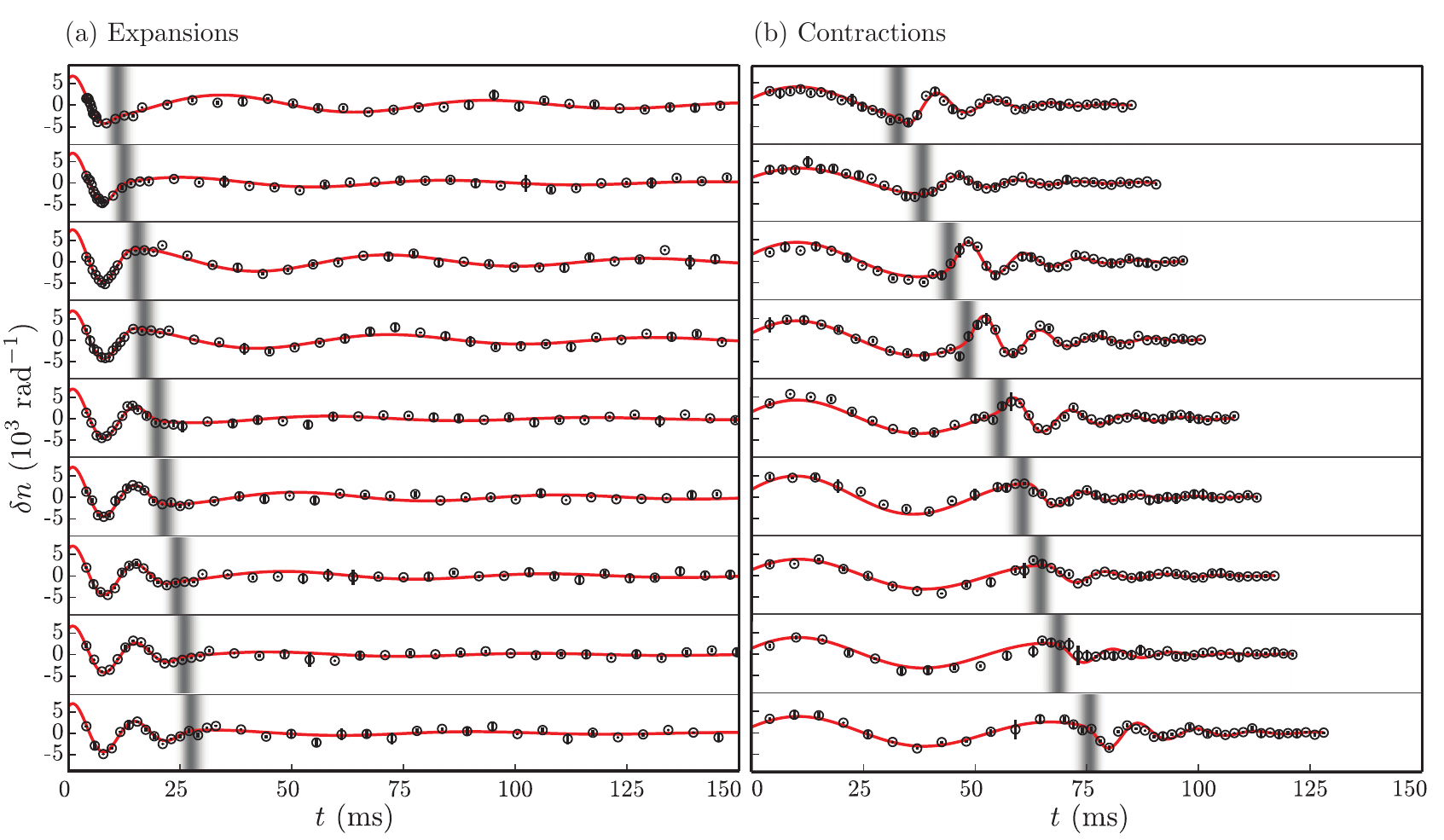}
    \caption{Phonon amplitude $\delta n$ as a function of time $t$ for (a) expanding and (b) contracting tori. 
    The symbols, curves, and grayscale bars are all as notated in Fig.~\ref{fig:ods}(d).
    The expansion data (a) used $R_{\rm i} = 11.9(2)~\mu{\rm m}$ and $R_{\rm f} = 38.4(6)~\mu{\rm m}$, and vice versa for contraction (b).
    $t_{\rm i}$ is varied from $6.5\ {\rm ms}$ to $23\ {\rm ms}$ for expansion and from $27\ {\rm ms}$ to $70\ {\rm ms}$ for contraction.
    Here, the red curves show simultaneous fits to \edit[our]{a} complete data set\edit[, as discussed in the text]{, which includes all expansions or contractions}.
    }
    \label{fig:datasets}
\end{figure*}

A\edit[n]{nearly-pure [$94(2)$~\%]} azimuthal phonon excitation with mode number $m$ = $1$ is generated by perturbing the toroidal BEC with a potential $V_{\rm ph}\sin(m\theta)$~\footnote{$V_{\rm ph}$ is set to 0.8 times the overall potential depth.}.
This repulsive potential---generated by the DMD---is applied for $2\ {\rm ms}$, imprinting the phonon's phase modulation onto the BEC.
After imprinting, the phonon evolves for an initial time $t_{\rm i}$ from $6.5\ {\rm ms}$ to $70\ {\rm ms}$, at which point the torus is expanded or contracted using an error function profile~\cite{Eckel2018}, with $10$~\%-$90$~\% rise time $3.6\ {\rm ms}$, and continues to evolve for up to $\approx 150\ {\rm ms}$.
For expansion, the initial and final radii are $R_{\rm i} = 11.9(2)\ \mu{\rm m}$ and $R_{\rm f} = 38.4(6)\ \mu{\rm m}$; these are reversed for contraction~\cite{ErrorFootnote}.
We detect the phonon at various points during the complete evolution using partial transfer absorption imaging (PTAI~\cite{Ramanathan2012}\edit[]{, with only one image per repetition of the experiment}) after a short $500\ \mu{\rm s}$ time of flight, giving the 2D density $n_{\rm 2D}(\rho,\theta)$ [see Fig.~\ref{fig:coordinates}].

The phonon excitation's density perturbation [see Fig.~\ref{fig:ods}(a)] is $\delta n_{\rm 2D} = n_{\rm 2D} - n^{0}_{\rm 2D}$, where $n^{0}_{\rm 2D}$ is the density with no phonon present. 
Integrating along $r$ gives the azimuthal density perturbation $\delta n_{\rm 1D}(\theta,t)$.
Figure~\ref{fig:ods}(b) shows the time evolution of $\delta n_{\rm 1D}$, and Fig.~\ref{fig:ods}(c) shows the resulting fit to Eq.~\eqref{eq:dynamics_DE}, from which we obtain both the red- or blueshift (via $c_{\theta}$) and the Hubble friction (from $\gamma_{\rm H}$).
In our system, the phenomenological damping $\gamma$ is observed to depend \edit{strongly} on radius\edit[, and we parameterize $\gamma$]{Instead, we parameterize the damping }in terms of the quality factor $Q=\omega/2\gamma$\edit[$ = c_\theta/2R\gamma$]{}, which \edit[eliminates most of the radial dependence present in $\gamma$]{is more independent of radius than $\gamma$} (see ~\cite{Marti2015,Kumar2016}).

Because the $3.6\ {\rm ms}$ expansion or contraction is a small fraction of the phonon oscillation period, the overall fit is insensitive to how $Q$ interpolates between $Q_{\rm i}$ to $Q_{\rm f}$. 
We therefore assume a simple linear dependence of $Q$ on $R$.
As shown in Fig.~\ref{fig:ods}(b), our data typically has less than one oscillation before $R$ changes; to reduce the uncertainty in $Q_{i}$ and $\omega(R_{\rm i})$, we include fixed-radius rings in a simultaneous fit.
These fits include as free parameters $\gamma_{\rm H}$, $Q_{\rm i}$, $Q_{\rm f}$, $\alpha$ as well as the \edit{initial} speed of sound $c_{\theta,{\rm i}}$, initial amplitude $\delta n_{\rm i}$, \edit{initial} temporal phase $\varphi_0$ \edit[]{, and an overall offset angle $\delta\theta$ capturing a small angular misalignment between the camera and DMD.} 
$c_\theta(t) = c_{\theta,{\rm i}}~(R(t)/R_{\rm i})^{-\alpha/2}$ follows the expected scaling.

Figure \ref{fig:ods}(d) summarizes the outcome of this fit.
The red curve is the time-dependent density perturbation $\delta n$ obtained from the full fit, while the circles plot $\delta n$ from independent fits to \edit[$\delta n \sin(\theta)$]{$\delta n \sin(\theta+\delta\theta)$} of each time-slice in Fig.~\ref{fig:ods}(b)\edit[.]{, thereby providing a 1D representation of the data in Fig.~\ref{fig:ods}(b).}
The blue curve displays the radius of the DMD pattern while the diamonds plot $R$ obtained from a 2D T-F fit to the observed density distribution
\footnote{By contrast with Ref.~\onlinecite{Eckel2018}, the BEC follows the contraction profile without overshoot or oscillation because of tighter radial confinement.}.
The gray band plots $\dot R/R$ during contraction, with maximum $\dot{R}/R\approx 1.53 \times \omega$.

We study the hypothesized impact of the phonon phase on the Hubble friction during expansion or contraction by changing $t_{\rm i}$ [see Fig.~\ref{fig:ods}(d)], thereby phase-shifting the phonon by $(c_{\theta,{\rm i}}/R_{\rm i}) t_{\rm i}$.
We define $t_{\text{peak}}$ as the time when the Hubble friction reaches its peak strength, i.e., when $|\dot{R}/R|$ is maximal.
The phase of the phonon at $t_{\text{peak}}$ is $\varphi_{\text{peak}} \equiv \int_0^{t_{\text{peak}}} dt ~ \omega (t)  +  \varphi_0$.
Figure~\ref{fig:datasets} shows example time-traces with multiple $t_{\rm i}$ for both expansion and contraction, providing a complete picture to investigate the strength of Hubble friction .
The black circles show the time evolution of the phonon amplitude $\delta n(t)$ for a range of $t_{\rm i}$ for both expansions (a) and contractions (b).

The red curves in Fig.~\ref{fig:datasets} show the results of global fits\edit[]{~\cite{Github_hubFriction2021}} of Eq.~\eqref{eq:dynamics_DE} to our complete dataset, which includes 17 contractions and 11 expansions. 
The parameters $\gamma_{\rm H}$, $\alpha$, $Q_{\rm i}$, $Q_{\rm f}$, $c_{\theta, {\rm i}}$ and $\delta n_{\rm i}$ are global, i.e., they are shared across all time traces.
For each time trace, $\delta n_{\rm i}$ is scaled by the atom number $N(t)$ for that trace, and $c_{\theta,{\rm i}}$ is correspondingly scaled by $\propto N(t) ^{\alpha/2}$~\footnote{This can be derived from Eq.(4.8) and Eq.(4.20) of \cite{Eckel2020}}; this accounts for both atom loss during and after expansion or contraction and for overall drifts in atom number during data acquisition.
Each global fit includes 7 additional time-traces, each with constant $R$, roughly from $R_{\rm i}$ to $R_{\rm f}$.
Because $c_\theta(R) \propto R^{\alpha/2}$ in stationary rings, these additional datasets further constrain $\alpha$.
We performed separate global fits for expansion and contraction data, giving an independent measure of their Hubble friction coefficients.
Finally, \edit[]{to mitigate potential systematic biases introduced by overfitting, } we perform these global fits in eight different ways, with the number of fit parameters varying between 32 and \edit[117]{101}\edit[]{ for $\approx 7.5\times 10^{4}$ independent data points}.
Each fit yields different best-fit values, but generally they agree within 2-$\sigma$.
These fitting methods differ on whether the temporal and azimuthal phases are shared across the time traces and if atom number varies within each time trace\edit[]{ (see supplemental material for details)}.
\edit[Due to the large number of data points, the degrees of freedom, in excess of  $1\times10^4$, do not vary significantly between the different methods.]{}
We take the mean of the values obtained from the eight methods as the best fit value. 
Their standard deviation is added in quadrature to the average 1-$\sigma$ uncertainty from the fit to obtain the final uncertainty in the measurement.

\begin{table}[b!]
\begin{tabular}{ c c c c c c c } 
\hline\hline
    & $Q_{\rm i}$  & $Q_{\rm f}$ & $\alpha$ &  $\gamma_{\rm H}$ & $c_{\theta, {\rm i}}$ & $\delta n_{\rm i}$ \\
    & & & & & (mm/s)  & (rad$^{-1}$) \\
\hline
Expansion & 3.5(1) &  4.4(2)  & 0.47(1) & 0.28(4) & 5.42(2) & 7.47(13) \\
Contraction & 7.8(3) & 3.5(1)  & 0.52(3) &  0.36(3) & 4.36(4) & 4.50(5) \\
\hline\hline
\end{tabular}
\caption{\label{tab:summary} Best fit global parameters.}
\end{table}

Table~\ref{tab:summary} lists the best-fit values, with $\gamma_{\rm H}$ different for contraction and expansion.
The values of $\alpha$ are in agreement with each other and are about $1/2$.
For our power-law potential model~\cite{Eckel2020}, $\alpha$ ranges from $1/2$ (for a harmonic potential) to $1$ (for a hard-wall potential).
Our average value of $\alpha\approx 0.495$ suggests that we have a harmonic potential in both $z$ and $r$.\edit[]{
The values of $c_{\theta, {\rm i}}$ and $\delta n_{\rm i}$ depend on the initial density, which is larger for expansions (i.e., smaller initial rings) than contractions.}
 
\begin{figure}[!t]
    \centering
    \includegraphics[width=8.5cm]{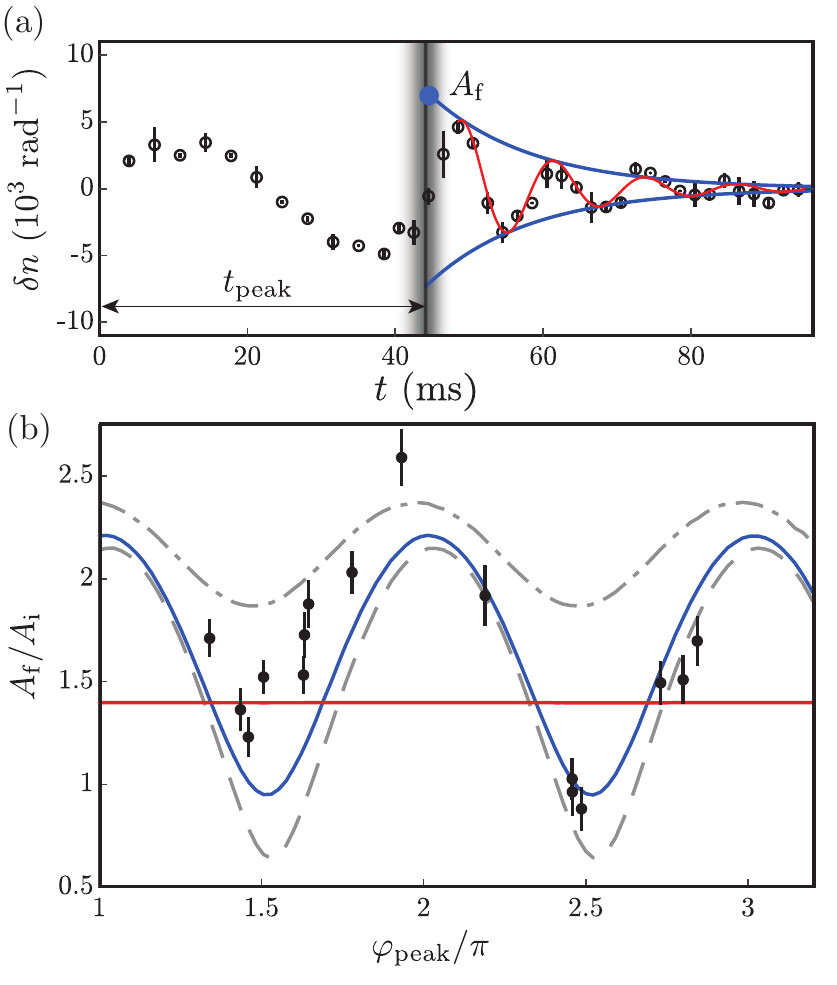}
    \caption{\label{fig:phase}
    Phonon amplitude vs. phase.
    (a) Data (black circles), fit (red curve), and oscillation envelope (blue curve) used to extract the amplitude $A_{\rm f}$ at $t_{\rm peak}$.
    The grayscale bar is as notated in Fig.~\ref{fig:ods}(d).
    (b) Ratio of amplitudes $A_{\rm f}/A_{\rm i}$ versus $\varphi_{\text{peak}}$, the oscillation's phase at $t_{\rm peak}$.
    The black circles plot the data\edit[.]{, and the error bars correspond to the fit uncertainty in the determination of $A_{\rm f}$.}
    The gray dashed, blue solid, and gray dashed-dot curves show the prediction of Eq.~\eqref{eq:dynamics_DE} for $\gamma_H=0$, 0.36, and $1$, respectively, with $\alpha=0.52$.
    The red line indicates the prediction for an adiabatic contraction.
    }
\end{figure}

Lastly, we confirm our expectation that the phonon phase $\varphi_{\text{peak}}$ has a marked impact on the amplitude following expansion or contraction in the non-adiabatic limit. 
Our experiments probe $1.3 \lesssim \varphi_{\rm peak}/\pi \lesssim 2.9$.
Fig.~\ref{fig:phase}(a) illustrates our process for obtaining the final amplitudes $A_{\rm f}$ where we fit the oscillatory behavior to an exponentially decaying sinusoid with the amplitude and temporal phase as free parameters (the remaining parameters are drawn from the global fits).
By contrast, the initial amplitude $A_{\rm i}$ is obtained from our global fit, from the envelope of the decaying sinusoid evaluated at $t_{\rm peak}$.
Figure \ref{fig:phase}(b) plots the fractional change in amplitude $A_{\rm f}/A_{\rm i}$ versus $\varphi_{\text{peak}}$ with black circles, and the solid blue curve depicts $A_{\rm f}$ obtained from our global fits \footnote{
The phenomenological damping term in Eq.~\eqref{eq:dynamics_DE} leads both $A_{\rm i}$ and $A_{\rm f}$ to decrease in a common-mode manner with increasing $\varphi_{\text{peak}}$, but that decrease is absent in the ratio $A_{\rm f}/A_{\rm i}$.}.
Our simulations (grey curves) show that the significant oscillations for $\gamma_{\rm H} = 0$, give way to more uniform gain with increasing $\gamma_{\rm H}$.
The measured values of $A_{\rm f}/A_{\rm i}$ are generally larger than would be expected for $\gamma_{\rm H} = 0$, showing Hubble amplification due to contraction.
Unlike Ref.~\onlinecite{Eckel2018}, which probed $1.8 \lesssim \varphi_{\rm peak}/\pi \lesssim 2.1$, where $A_{\rm f}/A_{\rm i}$ has little dependence on Hubble friction, our greater range of $\varphi_{\rm peak}$ allows us to better constrain $\gamma_H$.\editthird[]{
Using expansion and contraction in the same dataset provided a powerful tool for controlling systematic effects, resulting in increased precision and accuracy of $\gamma_H$.}
In addition to the overall oscillation, there appears to be some additional dependence on $\phi_{\rm peak}$ not captured by our model (data below $\phi_{\rm peak}/\pi<2$ generally lie above the $\gamma_H=0.36$ curve, and above for $\phi_{\rm peak}/\pi>2$).
This additional dependence may indicate a more complicated damping process for our phonons that could obscure our fitting for $\gamma_H$.

The observed oscillatory dependence of $A_{\rm f}$ on $\varphi_{\text{peak}}$ results from the rapid non-adiabatic, i.e. superluminal, contraction in this experiment.
The solid red curve emphasizes this point by plotting the simulated behavior for a slow adiabatic contraction, computed with $\gamma=0$.
No dependence on $\varphi_{\rm peak}$ is present in this limit, as the phonon would undergo many oscillations during expansion and therefore lose any dependence on initial phase.
The deviation from the adiabatic curve is associated with \edit{``classical'' stimulated emission or absorption (described by the mean field Gross-Pitaevskii equation, an interacting wave equation)} into or out of the phonon field, in much the same way that these processes have been observed in acoustic black holes~\cite{Steinhauer2014}.
Direct observation of spontaneous processes, i.e. pair production~\cite{Steinhauer2021}, would require an { increase} in our detection threshold.
While here we averaged three images per time point, the observation of spontaneous Hawking radiation in Ref.~\onlinecite{Kolobov2021} required a $\approx10^4$ image dataset. 

Our data generally agrees with the predictions of Refs.~\onlinecite{Llorente2019,Eckel2020}, with the notable exception $\gamma_{\rm H} \neq \alpha$.
\edit{
This discrepancy could be due to three possible effects.
First, the simple scaling of $c_\theta$ with $R$ holds only in the thin-ring approximation, and this can cause up to a 10~\% error in $\alpha$.
Second, while the excitation of higher azimuthal modes should have little impact (due to angular momentum conservation), expansion or contraction-driven mode mixing with higher excited radial modes (from non-adiabatic expansion) can contribute to an error in $\gamma_H$, which we estimate may be as high as 20~\%.
Third, we create large-amplitude phonons to maximize our detection signal, and this may lead  to non-linear damping effects~\cite{Katz2002}.
We note that general relativity is itself a non-linear wave equation, so related effects are potentially present in true cosmology.
In this linear regime, this complicates our measurement of $\gamma_{\rm H}$, and potentially causing the additional dependence on $\phi_{\rm peak}$ seen in Fig.~\ref{fig:phase}(b).}

For future experiments, our system is flexible enough to explore different metric scalings: to date we focused on quasi-one-dimensional universes, we could also potentially simulate two-dimensional (disc or square condensate) expansions or contractions where $\gamma_{\rm H}>1$, as suggested in Ref.~\onlinecite{Eckel2020}.
Our experimental setup could also readily explore other analogue gravity systems such as black hole horizons in 2D systems, where, for example the acoustic metric resulting from quantized vortices could open new directions~\cite{Coutant2015}.

\begin{acknowledgments}
The authors are grateful to T.~Jacobson for useful discussions and to S.~Mukherjee and M.~Doris for a careful reading of the manuscript.
This work was partially supported by NIST and NSF through the Physics Frontier Center at the Joint Quantum Institute. 
\end{acknowledgments} 

%

\begin{thebibliography}{35}
\bibitem [{Err()}]{ErrorFootnote}%
  \BibitemOpen
  \href@noop {} {}\bibinfo {note} {Unless stated otherwise, all uncertainties
  on numerical quantities are the uncorrelated combination of $1$-$\sigma$
  statistical and systematic uncertainties, while uncertainty bars in the
  figures include only the statistical contribution.}\BibitemShut {Stop}%
\end{thebibliography}
\end{document}